\documentclass[twoside]{article}
\usepackage{fleqn,espcrc2}
\input epsf
\input{epsf.sty}

\title{Melting of regular and decoupled vortex lattices in 
BSCCO crystals}

\author{M. Konczykowski$^{a}$, C. J. van der Beek$^{a}$,
M.V. Indenbom$^{a,b}$, E. Zeldov$^{c}$}

\address{Laboratoire des Solides Irradi\'{e}s, Ecole Polytechnique, 
91128 Palaiseau, France\\ 
\noindent $^{{\rm b}}$Institute of Solid State Physics, 142432
Chernogolovka, Moscow District, Russia\\
\noindent $^{{\rm c}}$Department of Condensed Matter Physics, Weizmann 
 Institute of Science, Rehovot, Israel\\}

\begin{document}

\begin{abstract}
The angular dependence of the first-order phase transition (FOT) in the 
vortex lattice in Bi$_{2}$Sr$_{2}$CaCu$_{2}$O$_{8}$ crystals was investigated by a
low frequency AC shielding technique (with the AC field $\parallel c$), in which the
static-field component parallel to $c$- ($H_{\perp}$) was varied with the 
in-plane field $H_{\parallel}$ held constant. 
The linear decrease of the FOT field $H_{\perp}^{FOT}$  with increasing $H_{\parallel}$ 
ends at a temperature--dependent critical value of $H_{\parallel}$. A new transition, 
marked by the abrupt drop of the $ab$-plane shielding current, appears at this 
point. We draw a new phase diagram with $H_{\parallel}$ and $H_{\perp}$ field components as 
coordinates; this features at least two distinct regions in the 
vortex solid phase, that are determined by the different
interplay between the pancake vortex-- and Josephson vortex lattice.
\vspace{1pc}
\end{abstract}

\maketitle


The use of the AC shielding technique for the identification of the 
first order transition (FOT) of the vortex lattice \cite{Morozov96} has allowed 
for the exploration of the phase diagram of Bi$_{2}$Sr$_{2}$CaCu$_{2}$O$_{8}$ 
(BSCCO) crystals in oblique fields. Early experiments have demonstrated that for large 
in-plane fields $H_{\parallel}$, the FOT occurs at a slightly lower 
value of the perpendicular ($\parallel c$) field  $H_{\perp}$ than 
predicted for either a purely 2D pancake vortex lattice, or for a 
highly anisotropic 3D vortex system\cite{Schmidt97}.
The linear dependence of the $c$-axis component of the FOT field  $H_{\perp}^{FOT}$
on $H_{\parallel}$ suggests that this is due 
to the partial suppression of $c$-axis phase coherence by
$H_{\parallel}$, and that an inclined field in fact 
penetrates as a combined lattice of pancakes and Josephson vortices 
\cite{Koshelev99}. Here, we present new measurements of 
the FOT in elevated parallel fields; the fact that one is dealing with 
a combined lattice inspired us to use the more straightforward 
experimental configuration in which $H_{\perp}$ and $H_{\parallel}$, 
and therefore the density of pancake vortices and Josephson vortices, 
are varied independently by simultaneous rotation of the magnet and adjustment 
of the total field magnitude. 




We have selected two BSCCO samples, cut from larger optimally and oxygen overdoped 
single crystals. After the samples' homogeneity was checked using
magneto-optical imaging of the flux penetration, a miniature 
($80 \times 80 \times 80$ $\mu$m$^{\rm 3}$ active volume) InSb Hall sensor 
was placed in the center of the sample top surface. The crystal 
together with the Hall sensor were placed in the centre of an excitation 
coil generating the low-frequency AC magnetic field,
mounted on the cold-finger of a pumped LN$_{2}$ dewar, and placed 
between the poles of a rotable electromagnet. In this set-up
the  AC excitation field, of amplitude $h_{ac}$ and frequency $f$, was 
always oriented along the crystal $c$-axis while the orientation and magnitude of the

\begin{figure}[h]
    \vspace{-14pt}
    \centerline{\epsfxsize 7cm \epsfbox{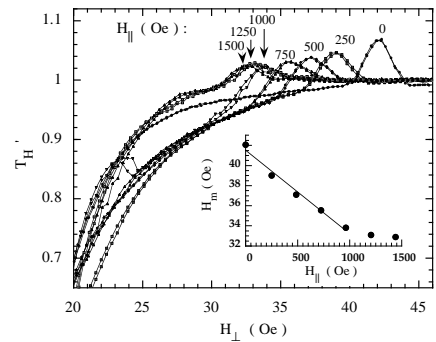}}
    \vspace{-20pt}
\caption{The $c$-axis field dependence of 
the in-phase transmittivity ($h_{ac}= 1$ Oe, $f = 7.75$ Hz) 
of the optimally doped sample at $T = 84$ K and various $ab$-plane fields. 
The paramagnetic peak identifies the
jump in the magnetization curve associated with the FOT. }
\label{fig:Fig1}
\end{figure}

\newpage

\noindent  DC magnetic  field were
independently controlled by a computer 
 system.
 
Two types of isothermal field-sweeps have been realized.  
First, $H_{perp}$ was varied with $H_{\parallel}$ held fixed; second, 
$H_{\parallel}$ was varied at fixed $H_{perp}$. The angular resolution 
of $10^{-3}$ $^{\circ}$ allowed scans of $H_{perp}$ in the 
range of 0 - 100 Oe on a $H_{\parallel}$--background of several kOe. 
The fundamental and third harmonic component of the AC magnetic 
induction were measured by two lock-in amplifiers. Typical variations of 
the normalized in-phase transmitivity $T_{H}^{\prime} $ are presented 
in Fig.~1. $T_{H}^{\prime}$ is defined as 
$(V^{\prime}-V^{\prime}_{0})/(V^{\prime}_{\infty}-V^{\prime}_{0})$, 
where $V^{\prime}$, $V^{\prime}_{\infty}$, $V^{\prime}_{0}$ denote the 
RMS in-phase AC Hall voltage, its value when
shielding can be neglected, and its value at full shielding.
We identify $H_{perp}^{FOT}$ by the paramagnetic peak 
in $T_{H}^{\prime}$, where this exceeds 1. This marks
the step in the DC-magnetisation curve\cite{Morozov96,Zeldov95}. 
The application of a moderate in-plane
field $\sim 50$ G yields a smaller $T_{H}^{\prime}$ for $ H_{\perp}  < 
H_{\perp}^{FOT}$, {\em i.e.} an enhancement of the shielding current. 
The linear decrease of $H_{perp}^{FOT}$ as function of $H_{\parallel}$ is consistent 
with earlier results \cite{Schmidt97}.

\begin{figure}[t]
\centerline{\epsfxsize 6.7cm \epsfbox{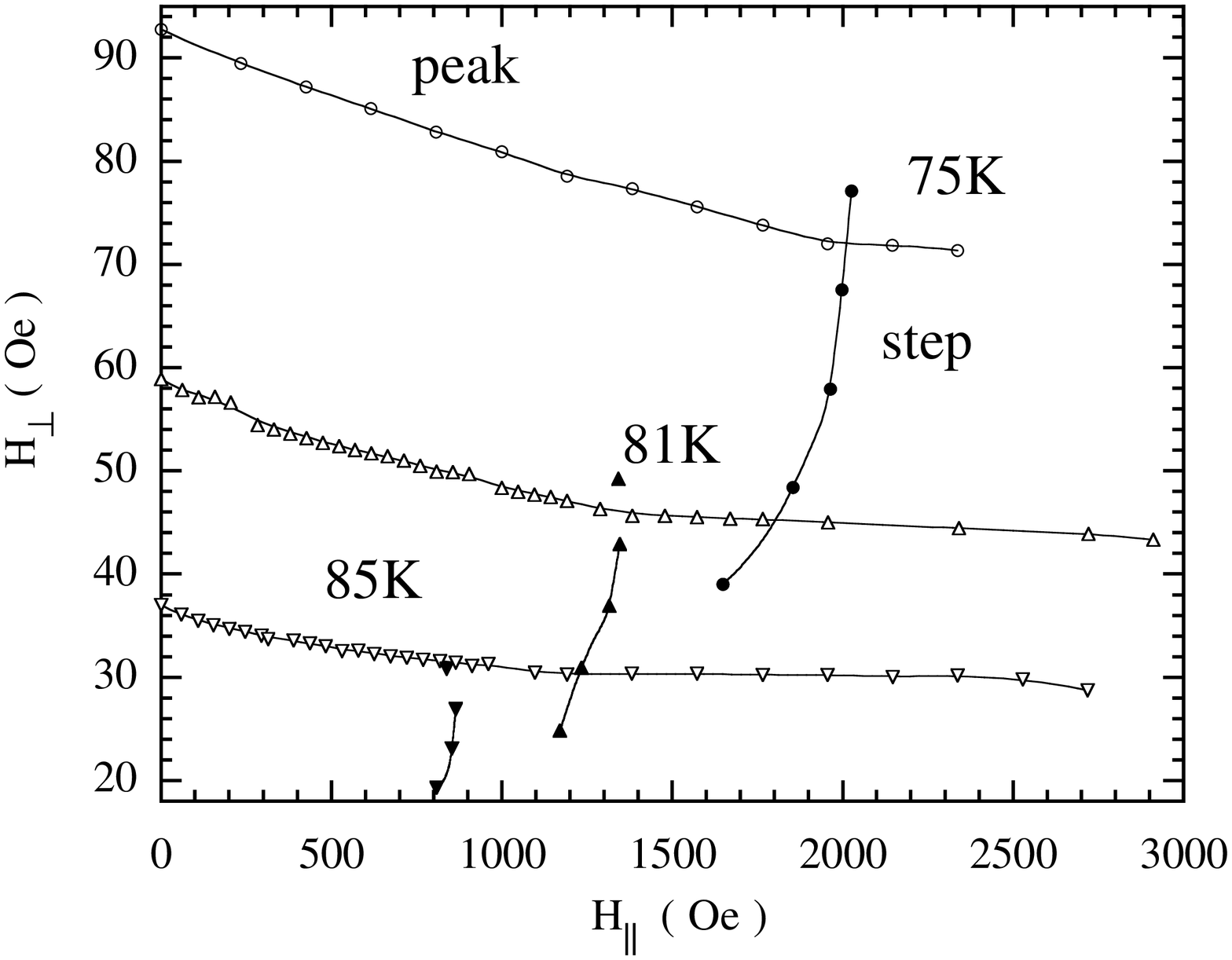}}
\vspace{-14pt}

\caption{Two field component phase diagram of the vortex lattice of BSCCO 
crystals, inferred from transmittivity measurements at various temperature.}
\label{fig:Fig2}
\vspace{-20pt}
\end{figure}

\noindent   
The novel result is the end of this decrease at high in-plane field.
In the optimally doped sample, we observe the paramagnetic 
peak in $T_{H}^{\prime}$ at the \em same \rm $H_{\perp}$ for all values of $H_{parallel} >  1$ kOe.
A new feature appears in the $T_{H}^{\prime}$ curves recorded with 
$H_{\parallel}$ slightly below 1 kOe: 
there is an additional $T_{H}^{\prime}$--minimum in the region of partial shielding 
($0 < T_{H}^{\prime} < 1$). The position of this minimum is independent 
of $h_{ac}$ and $f$, which indicates that
a transformation of the vortex lattice takes place at this point.
Other field scans in which $H_{\parallel}$ field was swept with 
$H_{\perp}$ held constant showed a
step-like drop of the shielding current at this new transition. Remarkably, 
the drop in the shielding current at this transition corresponds exactly to its enhancement 
observed after the initial application of a moderate in-plane field $\sim 50$ G. 
Using alternative $H_{\perp}$ and 
$H_{\parallel}$ scans we determine a two-field component phase diagram 
of the vortex lattice in oblique fields shown in Fig.~2. 


Three hypothetical transition lines can be drawn on the diagram presented in 
Fig.~2. In a highly anisotropic superconductor such as BSCCO, an 
oblique vortex lattice is expected to decompose into two superposed lattices of pancake vortices 
in the CuO$_{2}$ double layers and Josephson vortices  between layers. The enhancement
of the shielding current by a moderate $ab$-plane field may be 
considered the result of this decomposition. The melting line of the vortex 
lattice along the $c$- axis is expected 
to decrease with increasing $H_{\parallel}$ because of the 
introduction of Josephson vortices between the layers. This decrease 
saturates  at high $H_{\parallel}$ when the cores of the Josephson vortices 
overlap, or when the lattice of Josephson vorticres melts. Using a realistic value 
of the anisotropy factor $\gamma$, the line correpsonding to 
Josephson core overlap should be located at a much higher value of 
$H_{\parallel}$, of the order of 1 T. The invariance of the 
$H_{\perp}^{FOT}$ at high $H_{\parallel}$ indicates a uniform field distribution 
in the $ab$- direction, which is equivalent to the suppression of Josephson coupling. Thus the
quasi-vertical lines in Fig.~2 mark the crossover from the Josephson coupled regime 
to a regime where the planes are decoupled by the in-plane field.

\end{document}